\def\dover#1#2{\hbox{${{\displaystyle#1 \vphantom{(} }\over{
   \displaystyle #2 \vphantom{(} }}$}}
\def\teq#1{$\, #1\,$}                         % text equation
\def\erg{\varepsilon}         
\def\split{\gamma\to\gamma\gamma}               
{\catcode`\@=11                                                                 
\gdef\SchlangeUnter#1#2{\lower2pt\vbox{\baselineskip 0pt \lineskip0pt           
  \ialign{$\m@th#1\hfil##\hfil$\crcr#2\crcr\sim\crcr}}}}                        
\def\gtrsim{\mathrel{\mathpalette\SchlangeUnter>}}                              
\def\lesssim{\mathrel{\mathpalette\SchlangeUnter<}}    
\def\fsc{\alpha_{\hbox{\sevenrm f}}}
\def\pmb#1{\setbox0=\hbox{#1}%
  \kern-0.0125em\copy0\kern-\wd0
  \kern0.025em\copy0\kern-\wd0
  \kern-0.0125em\raise0.0433em\box0 }
\def\vol#1#2{$\;$ {\bf #1}, #2}                         % DO NOT DELETE %
\def\reference{\noindent\hangindent=.4truecm\hangafter=1}
\def\apj{{\it Ap. J.}}   \def\apjl{{\it Ap. J. (Lett.)}}

\def\apss{{\it Astr. Sp. Sci.}}   
\def\aap{{\it Astron. Astr.}}   
\def\aaps{{\it Astron. Astr. Supp.}}

\def\mnras{{\it Mon. Not. R. astr. Soc.}}

\documentstyle[twocolumn]{esa_sp_latex}

\begin{document}

% esa_sp_latex.tex
% PLEASE DO NOT EDIT THIS FILE. THANK YOU
%

\parindent 0pt
\parskip 10pt plus 1pt minus 1pt
\hoffset=-1.5truecm
\topmargin=-1.0cm
\textwidth 17.1truecm \columnsep 1truecm \columnseprule 0pt % FF

\input{psfig.tex}

\rm
\title{\bf THE ATTENUATION OF GAMMA-RAY EMISSION 
           IN STRONGLY-MAGNETIZED PULSARS}

\author{{\bf Matthew G.~Baring,}$^{1\hbox{\dag}}$\ {\bf Alice K.~Harding}$^1$ 
     {\bf and Peter L. Gonthier}$^2$ \vspace{2mm} \\
$^1$Lab. for High Energy Astrophysics, NASA Goddard Space Flight Center, 
    Greenbelt, MD 20771, U.S.A. \\
$^2$Deptartment of Physics, Hope College, Holland MI, U.S.A. \vspace{2mm} \\
\dag Compton Fellow, Universities Space Research Association}

\maketitle

\begin{abstract}

Gamma-rays from pulsars can be efficiently attenuated in their
magnetospheres via the mechanism of single-photon pair production and
also the exotic QED process of photon splitting, which become prolific
in fields approaching the quantum critical value of $B_{\rm
cr}=4.41\times 10^{13}$ Gauss.  Recently we have published results of
our modelling of strongly-magnetized $\gamma$-ray pulsars, which
focused on the escape or attenuation of photons emitted near the pole
at the neutron star surface in dipole fields, in a Schwarzschild
metric.  We found that pair production and splitting totally inhibit
emission above around 10--30 MeV in PSR1509-58, whose surface field is
inferred to be as high as $0.7B_{\rm cr}$.  Our model pulsar spectra
are consistent with the EGRET upper limits for PSR1509-58 for a wide
range of polar cap sizes.  Here we review the principal predictions of
our attenuation analysis, and identify how its powerful observational
diagnostic capabilities relate to current and future gamma-ray
experiments.  Diagnostics include the energy of the gamma-ray turnover
and the spectral polarization, which constrain the estimated polar cap
size and field strength, and can determine the relative strength of
splitting and pair creation.       \vspace {5pt} \\


\end{abstract}

\section{INTRODUCTION}
\label{sec:intro}

Magnetic one-photon pair production, \teq{\gamma\to e^+e^-}, has
traditionally been the only gamma-ray attenuation mechanism assumed to
operate in polar cap models for radio (e.g. Sturrock, 1971) and
gamma-ray pulsars (Daugherty \& Harding 1982, 1996; Sturner \& Dermer
1994).  Such an interaction can be prolific at pulsar field strengths,
specifically when the photons move at a substantial angle
\teq{\theta_{\rm kB}} to the local magnetic field.  Pair creation has a
threshold of \teq{2m_ec^2\approx 1.02}MeV for \teq{\theta_{\rm
kB}=90^\circ}.  The exotic higher-order QED process of the splitting of
photons in two, \teq{\split}, will also operate in the high field
regions near pulsar polar caps and until very recently, has not been
included in polar cap model calculations.  Magnetic photon splitting
has recently become of interest in neutron star models of soft gamma
repeaters (Baring 1995, Baring and Harding 1995, Thompson \& Duncan
1995), mainly because of their purportedly extreme fields (\teq{\gtrsim
10^{14}}Gauss).  Splitting becomes more effective in competition with
pair creation as a photon attenuation mechanism at higher field
strengths (Baring 1991). 

The key property of photon splitting that renders it relevant to
neutron star environs is that it has {\it no} threshold, and can
therefore attenuate photons below the threshold for pair production,
\teq{\gamma\to e^+e^-}.  Hence the importance of photon splitting in
gamma-ray pulsar models clearly needs to be assessed because (i) it is
an additional attenuation process for gamma-ray photons that can
produce cutoffs in the spectrum, and (ii) when it is comparable to pair
production, it will diminish the production of secondary electrons and
positrons in pair cascades while effectively softening the emission
spectrum.  Such ``quenching'' of pair creation can potentially provide
a pulsar ``death-line'' at high field strengths.  About a dozen radio
pulsars have magnetic fields, determined from dipole spin-down, above
$10^{13}$ Gauss.  This group includes PSR1509-58, the gamma-ray pulsar
having the lowest high-energy spectral turnover of $\sim 1$ MeV (Matz
et al. 1994, Wilson et al.  1993, Bennett et al.  1993).  Little
attention was paid to \teq{\split} in pulsar contexts prior to the
launch of the Compton Gamma-Ray Observatory (CGRO) in 1991 because
until then, the three known gamma-ray pulsars had estimated field
strengths of less than a few times \teq{10^{12}}Gauss.  The detection
of PSR1509-58 by the OSSE and Comptel experiments on CGRO provided the
impetus to focus on high-field neutron star systems.

Recent investigations of the role of \teq{\split} in gamma-ray pulsar
polar cap models have been performed by Harding, Baring and Gonthier
(1996, 1997) and Chang, Chen and Ho (1996).  In this paper, we outline
the importance of photon splitting for gamma-ray pulsar models,
focusing on the major features presented in Harding, Baring and
Gonthier (1997), and identifying powerful observational diagnostic
capabilities of our attenuation model and their relationship to current
and future gamma-ray experiments.  Principal diagnostics include the
maximum energy of gamma-ray emission, and the spectral shape and the
polarization below this maximum.  These constrain the estimated polar
cap size and/or field strength (i.e. location of emission on or above
the stellar surface), and can further determine the relative strength
of splitting and pair creation, which becomes salient for some subtle
physics issues pertaining to splitting.  As such, phase-resolved spectral
measurements in the soft gamma-ray band can provide a wealth of
diagnostic information, a goal that is readily achievable by the
Integral mission.  Future instrumentation with
polarization sensitivity will further enhance our understanding of
gamma-ray pulsars.

\section{SPECTRAL ATTENUATION IN GAMMA-RAY PULSARS}
\label{sec:atten}

Before discussing the attenuation of gamma-rays in pulsar
magnetospheres, it is instructive to briefly review photon splitting.
Since pair creation is widely invoked in pulsar models, it is
appropriate here to omit a detailed discussion of its properties,
referring the reader to Daugherty and Harding (1983).  While
\teq{\split} is forbidden in field free regions, it becomes quite
probable in neutron star fields, where \teq{B} becomes a significant
fraction of the quantum critical field \teq{B_{\rm cr}=m^2c^3/e\hbar
=4.413\times 10^{13}}Gauss.  Splitting is polarization-dependent in the
birefringent, magnetized vacuum, implying that polarized photons emerge
from an emission region.  The three polarization modes permitted by
CP-invariance in QED are \teq{\perp\to\parallel\parallel},
\teq{\perp\to\perp\perp} and \teq{\parallel\to\perp\parallel}.
However, Adler (1971; see also Usov and Shabad 1983) devised additional
restrictions, called polarization selection rules, by demanding absolute
four-momentum conservation {\it and} solving the dispersion relations
for photons in the polarized vacuum in the limit of weak dispersion.
This limits splitting to just one polarization mode
(\teq{\perp\to\parallel\parallel}) below pair creation threshold and
for \teq{B\lesssim B_{\rm cr}}.  Such selection rules are well-defined
only in cases of very weak linear dispersion, and could well be
modified in regimes of moderate or strong dispersion (\teq{B\gtrsim
B_{\rm cr}}), or by non-linear dispersion and field non-uniformity
effects; these modifications are not yet fully understood.

The study of the physics of \teq{\split} has at times had a tumultuous
history (Baring 1991, Harding, Baring and Gonthier 1997); the first
reliable calculations of its rate were performed in the early 70s (e.g.
Adler 1971, Papanyan and Ritus 1972), and splitting is still the
subject of some controversy.  For photon energies \teq{\erg  mc^2} with
\teq{\erg\ll 1}, and fields \teq{B\ll B_{\rm cr}}, the splitting rate
(e.g. Adler 1971) averaged over photon polarizations (e.g. Papanyan and
Ritus 1972; Baring 1991) can be expressed as an attenuation coefficient
\teq{T_{sp}}, which is the rate of splitting divided by \teq{c}:
\begin{equation}
   T_{sp} (\erg ) \approx {{\fsc^3}\over{10\pi^2}}\,         
   {\biggl({{19}\over{315}}\biggr) }^2\, {{mc}\over{\hbar}}\,\erg^5\,
   \biggl(\dover{B\sin\theta_{\rm kB}}{B_{\rm cr}} {\biggr)}^6\; ,
 \label{eq:splitrat}
\end{equation}
where \teq{\fsc\approx 1/137}, and \teq{\theta_{\rm kB}} is the angle
between the photon momentum and the magnetic field vectors; \teq{\erg}
is in units of \teq{mc^2}.  Note that \teq{\tau_{sp}(\erg )=T_{sp}(\erg
) R} is the optical depth for an emission region of size \teq{R}, when
B is spatially uniform.  Reducing \teq{\theta_{\rm kB}} or \teq{B}
dramatically increases the photon energy required for splitting to
operate in a neutron star environment, a property that also holds
for pair creation.  High field (\teq{B\gtrsim 0.3
B_{\rm cr}}) corrections to the above formula (e.g. see Harding, Baring
and Gonthier 1997, hereafter HBG97) for splitting diminish its
dependence on \teq{B}, causing the attenuation coefficient to saturate
above \teq{B\sim 4 B_{\rm cr}}.

In this paper, the attenuation of photons is determined as in HBG97 by
following photon paths in the dipole geometry of a neutron star field,
determining where they split or create pairs.  The photons originate at
different magnetic colatitudes $\theta$ on the neutron star surface and
propagate outwards, initially more-or-less parallel to the magnetic
field.  We choose the neutron star radius to be \teq{10^6}cm, and a
neutron star mass of \teq{1.4M_{\odot}}.  Opting for points of emission
on the stellar surface maximizes the average field strength along a
photon path, thereby producing the highest possible optical depths for
\teq{\split} and \teq{\gamma\to e^+e^-}.  Emission from above the
stellar surface, as would occur for processes such as curvature
radiation (e.g. Daugherty and Harding 1994) or magnetic Compton
upscattering (e.g. Sturner and Dermer 1994), pushes the threshold for
spectral opacity (i.e. the turnover) up in energy, since the rates for
splitting and pair creation are increasing functions of \teq{B} and
energy.  An important development in HBG97 was the inclusion of the
general relativistic effects of curved spacetime in a Schwarzschild
metric, following the treatment of Gonthier and Harding (1994).  These
included curved photon trajectories, affecting the angles photons make
to the field, the gravitational redshift of photon energy as a function
of distance above the neutron star surface, and an increase in the
dipole field strength (by about a factor of 1.4 at the pole) above 
flat spacetime values.  These effects all act to increase splitting and
pair creation optical depths and lower the maximum energies for
radiation transparency in the magnetosphere, typically by a factor of
2--3 below flat spacetime values.  Kerr metrics were not considered
since gamma-ray pulsar periods are much longer than their light
crossing times.

\subsection{Photon Escape Energies}

If a photon is attenuated via either absorption process after a
distance \teq{L} along its curved trajectory away from the pulsar
surface, then we call \teq{L} its attenuation length.  Clearly from the
behaviour in Eq.~(\ref{eq:splitrat}), and also for \teq{\gamma\to
e^+e^-}, attenuation lengths will be decreasing functions of the photon
energy \teq{\erg}; in fact, at high energies they vary as
\teq{\erg^{-5/7}} for splitting and \teq{\erg^{-1}} for pair creation
(HBG97).   Generally, they will be reached only after a photon has
propagated a sufficient distance to achieve a significant angle to the
field.  This criterion is easier to accomplish away from the magnetic
pole due to greater field curvature.  Hence attenuation lengths are
expected to be decreasing functions of the magnetic colatitude
\teq{\theta}; this is borne out in the detailed calculations of HBG97.
Since the rates of the two attenuation processes are strongly
increasing functions of energy, \teq{L} must approach infinity at some
finite energy for each process, below which photons are free to escape
the magnetosphere.  Such energies are called the escape energies, and
approximately delineate the energy at which spectral turnovers are
anticipated; they always exist due to the \teq{r^{-3}} decay of the
dipole field.

Fig.~1 illustrates escape energies, as determined in HBG97, for photons
initially propagating parallel to field lines (\teq{\theta_{\rm kB,0}=
0^\circ}) at the stellar surface.  These are strongly decreasing
functions of \teq{B} (roughly as \teq{B^{-6/5}} for splitting and
\teq{B^{-1}} for pair creation when \teq{B\ll B_{\rm cr}}) and
\teq{\theta} (\teq{\sim \theta^{-6/5}} for \teq{\split} and \teq{\sim
\theta^{-1}} for \teq{\gamma\to e^+e^-}).  These dependences and their
sensitivity are naturally expected to be borne out in spectral
turnovers from a population of sources; we argue below the value of
this strong diagnostic.  HBG97 showed that if the photons are permitted
to move at some small angle (at least around \teq{0.57^\circ}) to the
field initially, as might be the case for resonant Compton upscattering
polar cap models (e.g. Sturner and Dermer 1994, where particles with
Lorentz factors \teq{\gamma\gtrsim 100} emit the photons), the
sensitivity to colatitude is all but obliterated for
\teq{\theta\lesssim 2^{\circ}}.  Also evident in Fig.~1 is that for low
fields (below \teq{\sim 0.3B_{\rm cr}}), pair production escape
energies are below those for splitting, but in high fields, splitting
escape energies are lower at all $\theta$.  Hence, one expects photon
splitting is irrelevant to the consideration of the Crab and Vela
pulsars, whose spin-down fields are \teq{\sim 4\times 10^{12}}Gauss,
but would be very important for PSR1509-58, which has \teq{B\sim
3\times 10^{13}}Gauss.  Observe that the pair production escape
energies are bounded below by the pair threshold
$2mc^2/\sin\theta_{\rm kB}$, but photon splitting can attenuate photons
well below pair threshold.

%    FIGURE 1 GOES HERE
%
\vskip+0.0truecm
\centerline{\hskip 1.5truecm\psfig{figure=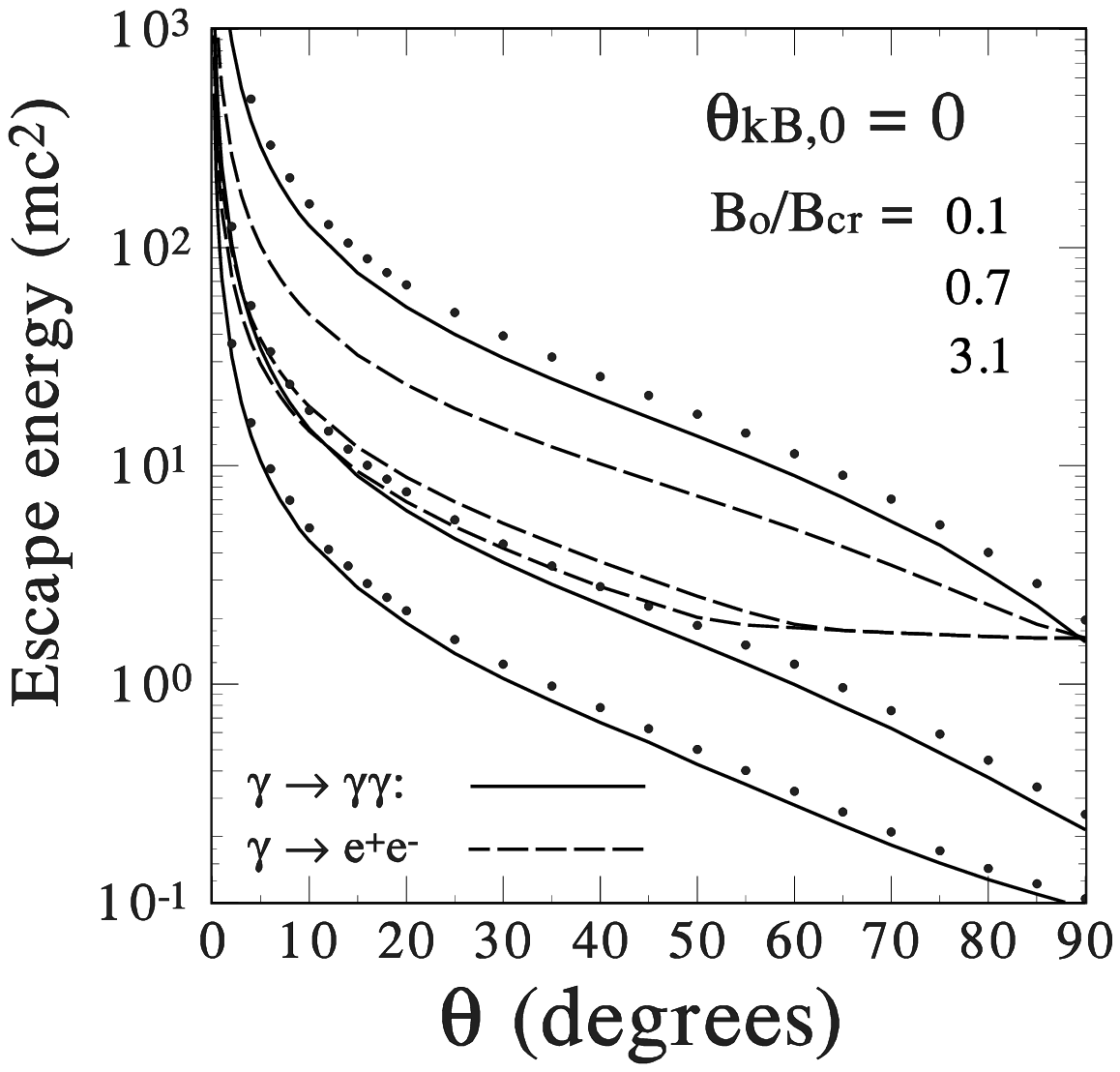,height=7.5truecm}}
\vskip-0.3truecm
{\smallskip \it Figure~1: The energy (in units of $mc^2$), below which
photons escape the magnetosphere without splitting (solid curves)
compared to the escape energies for one-photon pair production (dashed
curves) as a function of magnetic colatitude $\theta$ of the emission
point on the neutron star surface, for different surface dipole
magnetic field strengths (the escape energy drops as \teq{B}
increases).  The curves, which are for averages over the photon
polarizations, diverge near $\theta = 0$ because the photons are almost
parallel to the field lines throughout their path.  The dots depict the
escape energies for the polarization mode
\teq{\perp\to\parallel\parallel}.    \smallskip}

\subsection{Spectral Attenuation in PSR1509-58}

Here, the depiction of spectral attenuation caused by splitting and
pair creation will be focused on the case of the high field gamma-ray
pulsar PSR1509-58.  This is because the ``GeV'' cutoffs and
reprocessing in the Crab and Vela spectra are well-studied (e.g.
Daugherty and Harding 1982, 1996), and as mentioned above, photon
splitting is unimportant in these two sources.  We adopt the high
revised spin-down estimate \teq{B=0.7B_{\rm cr}} for PSR1509-58
following Usov and Melrose (1995).  Photons are injected at the stellar
surface with a canonical unpolarized power-law continuum that is chosen
to be consistent with the OSSE data points (e.g. see Matz et al. 1993),
and we determine emergent spectra ignoring (for simplicity) any photon
generation by created pairs.  Details of pair cascading will be the
subject of future work.  Fig.~2 shows the differential energy spectrum
obtained in the case where both photon polarizations produce pairs, but
only one splits (\teq{\perp\to\parallel\parallel}) according to
Adler's (1971) selection rules, so that no cascading
of photons ensues.  Strong polarization of the continuum results from
unpolarized injection; clearly a polarized injection could be either
further polarized, or depolarized by the attenuation.  The spectra are
determined for injection of photons at a small angle to the surface
field.  This introduces a dependence of the energy of the cutoff on the
polarization state of the photon, which arises because of the different
cutoff energies for photon splitting and pair creation.  Such a
property, which can be a powerful observational diagnostic, all but
disappears when the photons start off parallel to the field.  Note,
however, that polarization of the continuum below the cutoff is still
present when \teq{\theta_{\rm kB,0}= 0^\circ}.

%    FIGURE 2 GOES HERE
%
\vskip-0.1truecm
\centerline{\hskip 0.8truecm\psfig{figure=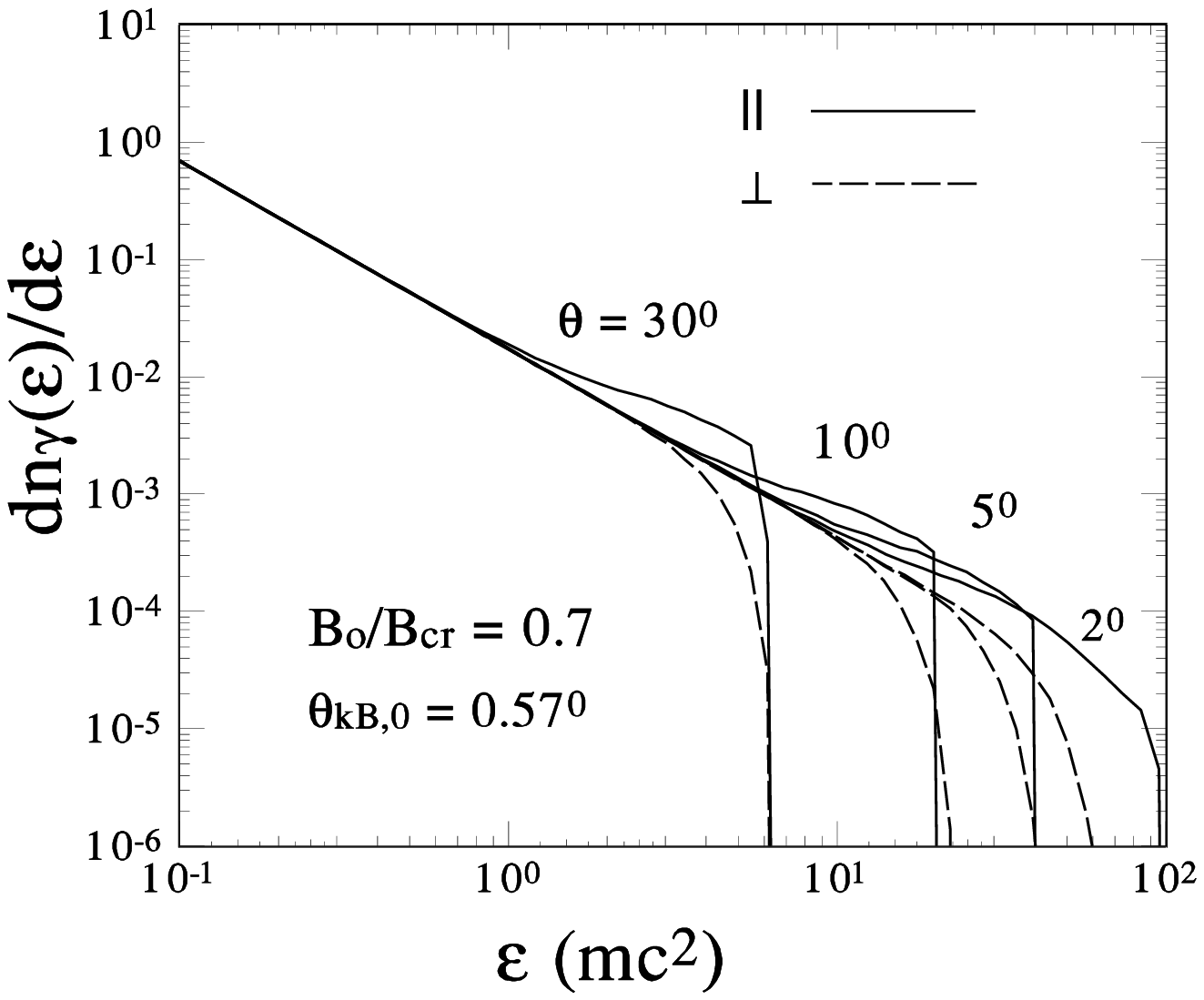,height=7.4truecm}}
\vskip-0.3truecm
{\smallskip \it Figure~2: Polarized spectra (see HBG97) for partial
photon splitting cascades, assuming unpolarized power-law emission (of
index \teq{\alpha =1.6}, the OSSE best-fit value) not parallel to the
magnetic field ($\theta_{\rm kB,0} = 0.57^\circ$), at different
magnetic colatitudes, $\theta$, as labelled.  Here only photons of
polarization \teq{\perp} split, while those of either polarization
produce pairs.  The normalization of the spectrum is arbitrary.
\smallskip}

%    FIGURE 3 GOES HERE
%
\vskip-0.1truecm
\centerline{\hskip 0.6truecm\psfig{figure=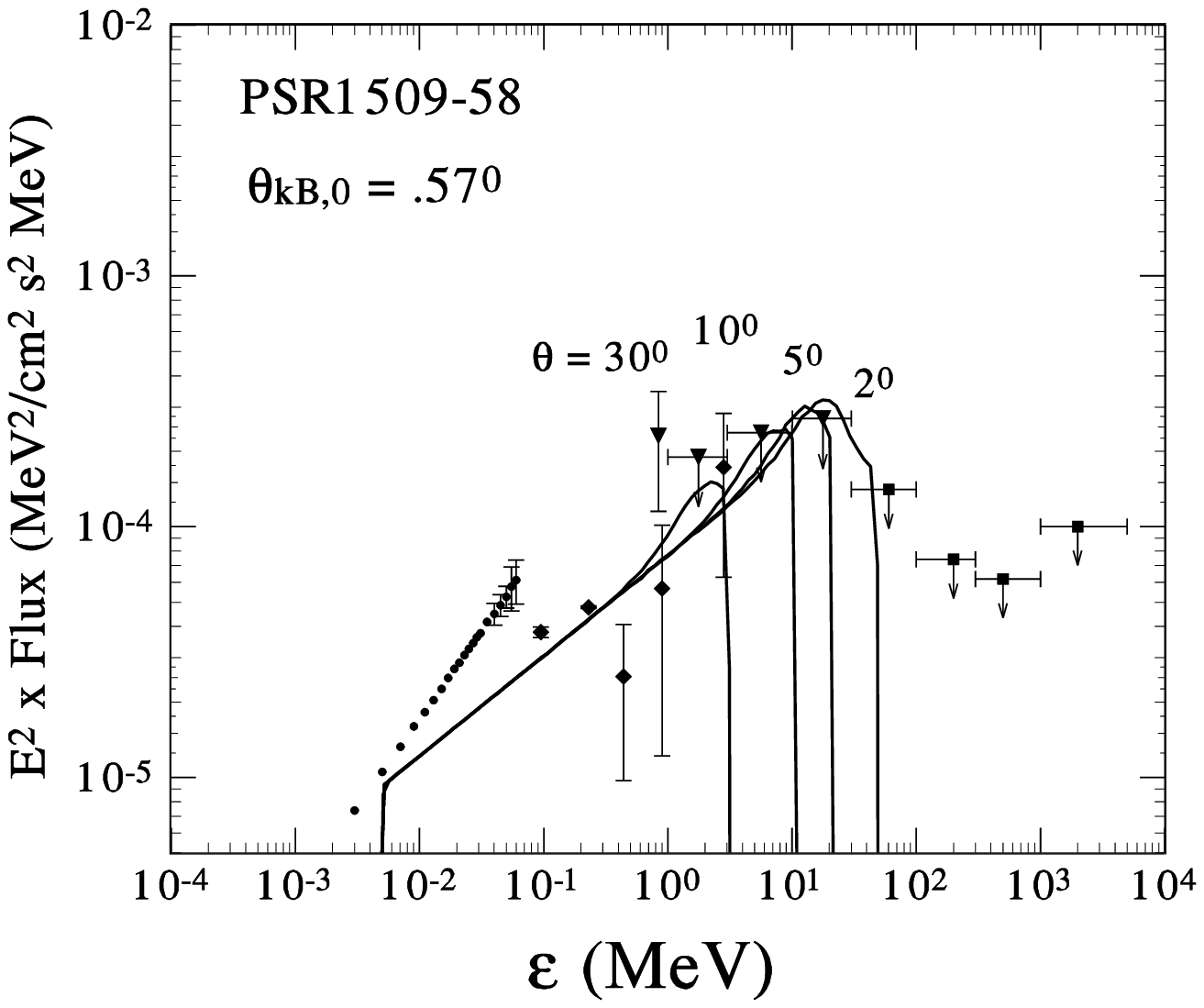,height=7.0truecm}}
\vskip-0.3truecm
{\smallskip \it Figure~3:  The polarization-averaged spectrum,
multiplied by energy squared, for different colatitudes \teq{\theta} of
surface emission, with only one mode of splitting permitted as in
Fig.~2.  The data points were obtained by the Ginga, Einstein and OSSE
instruments, while upper bounds at higher energies belong to Comptel
and EGRET viewings (see HBG97 for references).   \smallskip}

Fig.~3 depicts this spectrum in a ``\teq{\nu}-\teq{F_{\nu}}''
representation, along with observational data and upper limits.  A wide
range (\teq{2^\circ\lesssim\theta\lesssim 25^\circ}) of polar cap sizes
match the observations for the chosen field strength.  Restricting
splitting to just one polarization mode
(\teq{\perp\to\parallel\parallel}) below pair creation threshold may be
relaxed if modifications to Adler's (1971) polarization selection rules
prove significant, as discussed above.  If this arises (e.g. through
microscopic momentum non-conservation), and permits all three modes of
splitting to proceed, a full cascade ensues with several generations of
splitting and polarization state switching, as described by Baring
(1995).  This pushes the photon spectrum to lower energies since photon
splitting tends to dominate pair creation at such high field
strengths.  Large spectral peaks will result, with strong polarization
signatures, as is illustrated in HBG97, and the phase space for polar
cap sizes permitted by the observational data diminishes.  The
concurrent enhanced quenching of pair creation would then be very
important for radio and gamma-ray pulsar models.

\section{DISCUSSION}

In the light of the attenuation results presented here, the diagnostic
capabilites of future experiments appear great, and will significantly
impact our understanding of gamma-ray pulsars.  In observations of
individual sources, if polarization measurements are not possible, it
will be difficult to pin down the field strength and polar cap size
separately; the energy of the cutoff clearly depends on both these
quantities.  However, if polarization detections yield information such
as depicted in Fig.~2, where it becomes possible to observationally
determine the ratio of escape energies for splitting and pair creation,
then \teq{B} and the cap size for emission at any given height above
the stellar surface can both be constrained at the same time.  The
spectral shape below the cutoff is a strong function of the number of
modes of splitting that operate in the system.  Prominent bumps should
appear in source spectra (HBG97) unless the polarization selection
rules that are mentioned above restrict photon splitting to just the
\teq{\perp\to\parallel\parallel} mode.  Hence just through a
determination of spectral shape, Integral and other gamma-ray
instruments may well be able to discriminate between the operation or
otherwise of the selection rules in pulsar magnetospheres, thereby
impacting our understanding of this aspect of strong field QED, a very
enticing prospect.

As a dramatic increase in the number of observed gamma-ray pulsars in
the not too distant future is anticipated, we note diagnostic advances
can be made with populations of these sources.  Firstly, the polar cap
scenario considered here makes a definitive prediction that the
spectral cutoff energy decreases quickly with increasing spin-down
field strength.  While the Crab, Vela and PSR1509-58 provide a range of
field strengths that suggests this behaviour, it is important to
confirm or deny this trend with other sources.  The recent marginal
detection of PSR0656+14, whose spin-down field is \teq{\sim
10^{13}}Gauss, at around 100 MeV (Ramanamurthy et al. 1996) conforms to
this trend.  A much larger population of detected highly-magnetized
gamma-ray pulsars will probe this hypothesis and probably provide the
capability to discriminate between the polar cap and outer gap models,
since the outer gap scenarios (e.g. Romani and Yadigaroglu, 1995) are
unlikely to produce this correlation with field strength.  If this
trend is confirmed, we may further have the potential to discern
between the physical mechanisms responsible for the gamma-ray
emission.  Resonant Compton upscattering and curvature radiation
produce different angular beaming patterns of radiation, and therefore
a different spread of cutoff energies for given \teq{B} and
distributions of polar cap sizes.  Hence distributions of key spectral
parameters can refine our modelling of these sources.  Furthermore, the
future full treatment of pair cascades will reveal the relative
importance of curvature (flat) and synchrotron (steep) contributions to
the gamma-ray spectra of strongly-magnetized pulsars.  This will define
a correlation between source spectral index and \teq{B}, and therefore
cutoff energy, trends that can be verified or refuted observationally.
In summary, clearly the era of the Integral mission will significantly
advance our understanding of gamma-ray pulsars.

%\section*{ACKNOWLEDGMENTS}

%\begin{thebibliography}{}

%\end{thebibliography}

\clearpage

% Have here your figure (s) included as "file" in the \epsfig statement
% Have one \begin ...\end per figure
% Of course this figure can be inserted somewhere in the text....

%\begin{figure}[p]
%  \begin{center}
%    \leavevmode
%\epsfig{file=post1.eps, width=9.0cm, bbllx=0pt, bblly=260pt,
%  bburx=624pt, bbury=547pt, clip=}
%  \end{center}
%  \caption{\em The performance of the baseline configuration for the
%    space-based echelle measurements}
%  \label{fig:echbasegaia}
%\end{figure}

\end{document}